\documentclass[prb,twocolumn,showpacs,superscriptaddress,amsmath,amssymb,floatfix]{revtex4-1}
\usepackage{graphicx}
\usepackage{dcolumn}
\usepackage{bm}
\usepackage{amssymb}
\usepackage{epsfig}
\usepackage{epsfig}
\usepackage{amsmath}
\usepackage{amssymb}
\usepackage{amsfonts}
\usepackage{mathptmx}
\usepackage{dcolumn}
\usepackage{eucal}
\usepackage{epstopdf}
\usepackage{bm}
\usepackage{color}
\usepackage{graphicx}
\usepackage{natbib}
\usepackage{comment}
\usepackage[colorlinks,linkcolor=blue,citecolor=blue]{hyperref}

\specialcomment{unmodified}{\begingroup\color{blue}}{\endgroup}
\specialcomment{AB}{\begingroup\color{red}}{\endgroup}

\begin{document}
\title{Ferromagnetic insulator-based superconducting junctions as sensitive electron thermometers}
\author{F. Giazotto}
\email{francesco.giazotto@sns.it}
\affiliation{NEST Istituto Nanoscienze-CNR  and Scuola Normale Superiore, I-56127 Pisa, Italy}
\author{P. Solinas}
\affiliation{SPIN-CNR, Via Dodecaneso 33, 16146 Genova, Italy}
\author{A. Braggio}
\affiliation{SPIN-CNR, Via Dodecaneso 33, 16146 Genova, Italy}
\affiliation{I.N.F.N. Sezione di Genova Via Dodecaneso 33, 16146, Genova, Italy}
\author{F. S. Bergeret}
\email{sebastian\_bergeret@ehu.es}
\affiliation{Centro de F\'{i}sica de Materiales (CFM-MPC), Centro Mixto CSIC-UPV/EHU, Manuel de Lardizabal 4, E-20018 San Sebasti\'{a}n, Spain}
\affiliation{Donostia International Physics Center (DIPC), Manuel de Lardizabal 5, E-20018 San Sebasti\'{a}n, Spain}
\begin{abstract}
We present an exhaustive theoretical  analysis of charge and thermoelectric transport in a normal metal- ferromagnetic insulator-superconductor (NFIS) junction, and explore the possibility of its use as a sensitive thermometer. We investigated the transfer functions and the intrinsic noise performance for different measurement configurations. A common feature of all configurations is that the best temperature noise performance is obtained in the non-linear temperature regime for a structure based on an europium chalcogenide ferromagnetic insulator in contact with a superconducting Al film structure. 
For an open-circuit configuration, although the maximal intrinsic temperature sensitivity can achieve $10$nKHz$^{-1/2}$, a realistic amplifying chain will reduce the sensitivity up to $10$$\mu$KHz$^{-1/2}$. 
To overcome this limitation we propose a measurement scheme in a closed-circuit configuration based on state-of-art SQUID detection technology in an inductive setup. In such a case we show that temperature noise can be as low as  $35$nK Hz$^{-1/2}$. 
We also discuss a temperature-to-frequency converter where the obtained thermo-voltage developed over a Josephson junction operated in the dissipative regime is converted into a high-frequency signal. We predict that the structure can generate frequencies up to  $\sim 120$GHz, and transfer functions up to $200$GHz/K at around $\sim 1$K. If operated as electron thermometer, the device may provide temperature noise lower than $35$nK Hz$^{-1/2}$ thereby being potentially attractive for radiation sensing applications.
\end{abstract}
\maketitle

\section{Introduction}

Recent theories have shown that the spin-splitting induced in a superconductor (S) placed in contact with a ferromagnetic insulator (FI) can be exploited in different kinds of spin caloritronic devices such as heat valves~\cite{valve,longpaper} or thermoelectric elements~\cite{Machon,Ozaeta,Machon2,beckmann}. They can be used as building blocks in phase-coherent thermoelectric transistors~\cite{transistor}, and for the creation of magnetic fields induced by a temperature gradient in Josephson junctions (JJs) due to the thermophase effect \cite{thermophase}.  NFIS junctions have been also proposed  for efficient electron cooling \cite{giazottormp2006} of the  normal metal N~\cite{kawabata}.  The possible applications of superconductor-ferromagnetic structures for  thermoelectrics has been also highlighted in a recent review article.\cite{Linder2015}
\begin{figure}[t!]
\includegraphics[width=\columnwidth]{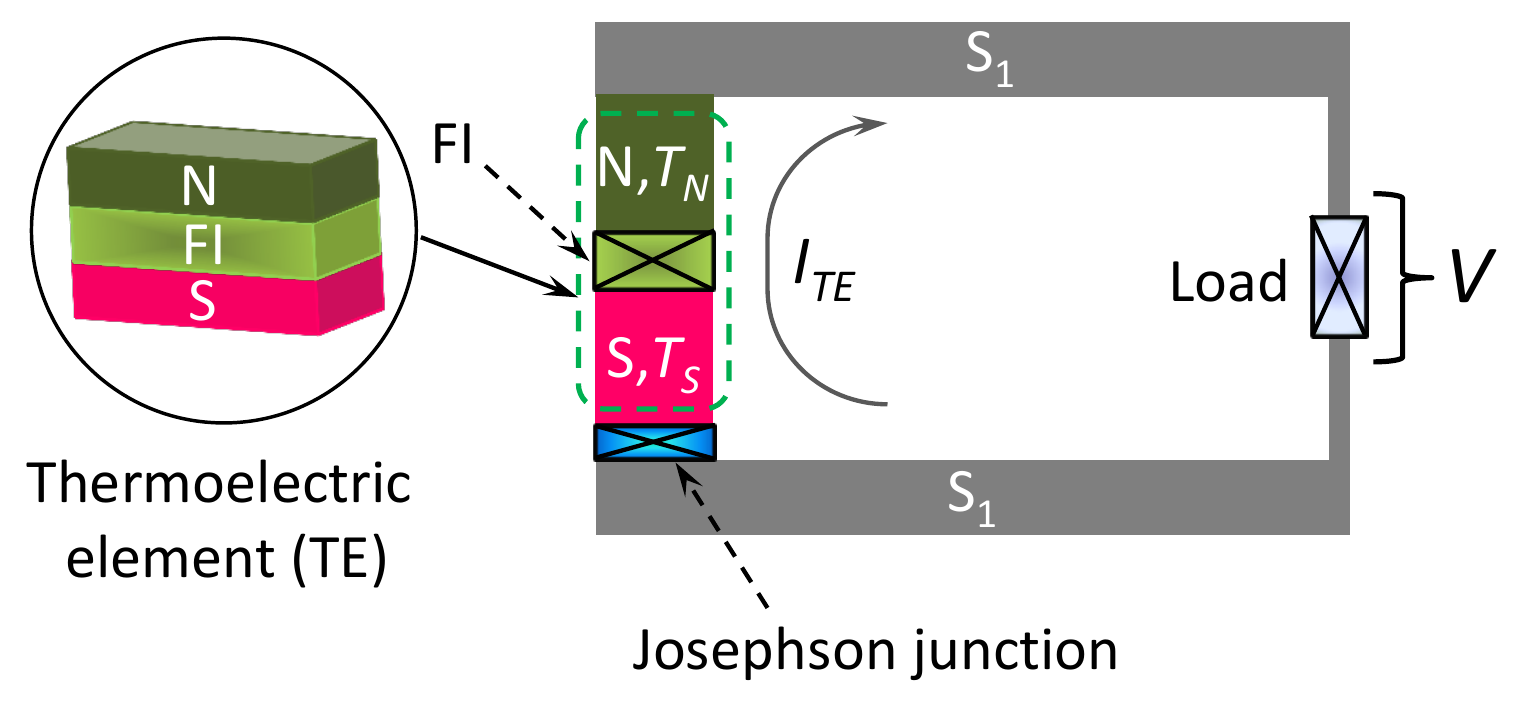}
\caption{\label{fig1} 
(Color online) General scheme of the device based on  a normal metal-ferromagnetic insulator-superconductor (NFIS) junction. The latter is shown in the blow-up as stacked layers of different materials.
$T_S$ and $T_N$ denote the temperature in S and N, respectively, $I_{TE}$ is the thermocurrent circulating in the circuit, and $V$ is the thermovoltage developed across the thermoelectric element (TE). S$_1$ is a superconductor contacted to both ends of the TE which contains a load with resistance $R_L$. The latter is intended to be an open-circuit $R_L\to\infty$, a dissipationless closed-circuit $R_L=0$, or in the form of a generic Josephson element operated in the dissipative regime in order to convert the thermovoltage $V$ into radiation at the Josephson frequency.}
\end{figure}

In the present work we  theoretically analyze charge and thermoelectric transport in a prototype structure  based on the FIS building block, and explore its application as an ultra-sensitive electron 
thermometer~\cite{schmidt2003,gasp2012,torresani2013,faivre2014,gasparinetti2015,giazottoPJS,mottonen2014,faivre2015} and, eventually as a temperature-to-frequency converter.  
Our system consists of a normal metal-ferromagnetic insulator-superconductor (NFIS) junction, denoted here as the thermoelectric element  (TE), which is connected, via the superconducting wires S$_1$, to a generic load element, as shown in Fig. \ref{fig1}. A temperature difference localized between the N and S side of the TE induces a thermoelectric signal\cite{Ozaeta}. We consider three different configurations of the load resistance $R_L=\infty$ (open circuit), $R_L=0$ (closed circuit) and finite load $R_L=R_{JJ}$ where we close the system over a generic Josephson element, in the dissipative regime, with shunting resistance $R_{JJ}$. Depending on the configuration the device  will operate in different regimes: i) \emph{Seebeck regime}, where a Seebeck \emph{thermovoltage} ($V$) is generated across the TE element at open-circuit; ii) \emph{Peltier regime}, where the gradient of temperature generates a circulating \emph{thermocurrent} that can be probed by an inductive measurement for closed-circuit.  Here we explore  both regimes that includes 
 an estimate of the intrinsic noise and the best expected temperature sensitivity with state-of-art technology for  signal detection. We discuss the advantages and the drawbacks of the different configurations and  show that operated within  the  non-linear regime,  the intrinsic noise of the device is  reduced. In particular, our numerical results show that the noise performance is mainly determined by  the junction differential resistance $R_d$ which can be drastically reduced beyond  the linear-response   regime with respect to the temperature.  We finally discuss how the generated thermovoltage can  induce an ac-Josephson effect with a supercurrent oscillating at a frequency $\nu=|V|/\Phi_0$ \cite{barone}, where $\Phi_0\simeq 2.067\times 10^{-15}$ Wb is the flux quantum.  The frequency $\nu$ can be measured with great accuracy providing accurate and fast  information about temperature difference across the TE.

The paper is organized as follows: In Sec.~\ref{model} we briefly present the general formalism and the  expressions for the  electric current flowing  through the NFIS junction and the noise  as a function of  all the parameters involved in the system. 
With the help of this expression we analyze in Sec.~\ref{seccurrent} the electric and thermoelectric response of the TE in the non-linear response regime. In particular, we show the impact of the exchange field as well as the role of the barrier polarization on the charge current. In Sec.~\ref{measurement} we discuss the different measurement configurations of the device analizing the effect of the load resistance $R_L$ over the thermo-electrical properties of TE recalling the results for the linear regime in Sec.~\ref{linear}. The evaluation of the intrinsic noise properties of the NFIS junction is done both for the linear an non-linear regime. Assuming a realistic device based on europium sulfide (EuS) as FI and superconducting aluminum (Al), operating at low temperatures, we discuss the open-circuit and closed-circuit configurations respectively in Sec.~\ref{opencircuit} and Sec.~\ref{closedcircuit}. In those sections we also discuss the temperature noise performance taking into account the most simple measurement scheme with actual state-of-art technologies. 
Finally, in Sec.~\ref{temptofreq}, we discuss the temperature-to-frequency conversion scheme where the thermovoltage developed across the NFIS junction is converted into a high-frequency signal by a Josephson element driven into the dissipative regime. The full temperature-to-frequency conversion capability of the NFIS junction is analyzed, investigating as well the  temperature noise performances.
We summarize our results in Sec. \ref{summary}.
\begin{figure}[t!]
\includegraphics[width=\columnwidth]{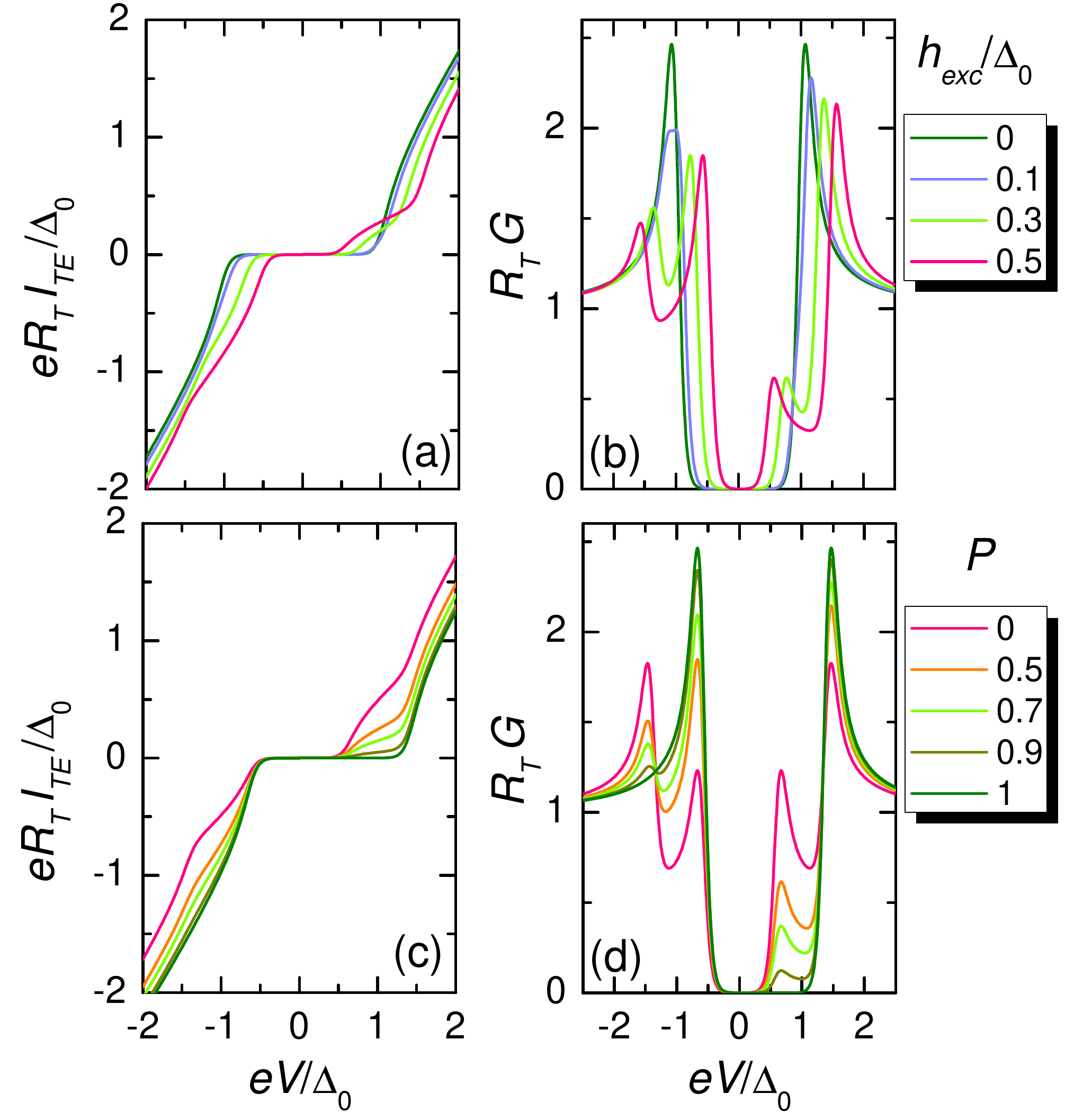}
\caption{ 
(Color online) {\bf Characterization of the TE for $\mathbf{T_S=T_N}$} (a) Current vs voltage ($I_{TE}-V$) characteristics of the TE element calculated at $T_S=T_N=0.1 T_c$, $P=0.5$ and for a few values of $h_{exc}$. (b) Differential conductance vs voltage ($G-V$) characteristics of  TE calculated for the same parameters as in panel (a).
(c) $I_{TE}-V$ and (d) $G-V$ characteristics of TE calculated at $T_S=T_N=0.1 T_c$, $h_{exc}=0.4\Delta_0$ and for a few values of $P$. 
$\Delta_0=1.764k_BT_c$ is the zero-temperature, zero-exchange field superconducting gap, $T_c$ denotes the critical temperature, and $R_T$ is the normal-state resistance of TE.
}
\label{fig2}
\end{figure}

\section{Model}
\label{model}
It is instructive to start with the description of  the NFIS building block. The interaction between the spin of the conducting electrons in S and the localized magnetic moments in FI lead to an effective exchange interaction in  S  that decays over  the superconducting coherence length $\xi_0$\cite{Tokuyasu1988}. 
We assume that the S layer is  thinner than  $\xi_0$, so that the exchange field ($h_{exc}$) induced in S by FI is spatially homogenous. 
In such a case the superconductor density of the states (DoSs) is given by   the sum of the densities for spin-up ($\uparrow$) and spin-down ($\downarrow$) quasiparticles, 
\begin{equation}
N_{\uparrow,\downarrow}(E)=\frac{1}{2}\left|{\rm Re}\left[\frac{E+i\Gamma \pm h_{exc}}{\sqrt{(E+i\Gamma \pm h_{exc})^2-\Delta^2}}\right]\right|.
\end{equation}
Here $\Delta(T_S,h_{exc})$ is the  pairing potential that depends both on temperature $T_S$ in S and $h_{exc}$, and it is  computed self-consistently in a  standard way \cite{tinkham} from the gap equation
\begin{equation}
\textrm{ln}\left(\frac{\Delta_0}{\Delta}\right)=\int_{0}^{\hbar \omega_D}dE\frac{f_+(E)+f_-(E)}{\sqrt{E^2+\Delta^2}},
\end{equation}
where $f_{\pm}(E)=\left\{1+\textrm{exp}[\frac{1}{k_B T}(\sqrt{E^2+\Delta^2}\mp h_{exc})]\right\}^{-1}$, $\omega _D$ is the Debye frequency of the superconductor, $\Delta_0$ is the zero-temperature, zero-exchange field superconducting pairing potential, and $k_B$ is the Boltzmann constant.
Furthermore, $\Gamma$ accounts for broadening, and for an ideal superconductor $\Gamma\rightarrow 0^+$ \cite{dynes}. 

We are interested in the current  through the NFIS junction which in the tunneling limit considered here  is given  by \cite{Ozaeta}
 \begin{equation}
 \label{current_TE}
 I_{TE}=\frac{1}{eR_T}\int_{-\infty}^{\infty} dE\left[N_++PN_-\right]\left[f_N(V,T_N)-f_S(T_S)\right]\; .
 \end{equation}
Here $R_T$ is the normal-state resistance of the tunneling junction and  $N_\pm=(N_\uparrow\pm N_\downarrow)$.  Notice that in the tunneling limit the Andreev reflection is negligible small and hence no superconducting proximity effect in N takes place. We assume thermalization on both, the S and the N layer neglecting any deviation of the distribution functions from their equilibrium form \cite{Silaev2015}: $f_{S}(T_S)=[1+\textrm{exp}(E/k_BT_S)]^{-1}$ and $f_{N}(V,T_N)=[1+\textrm{exp}[(E+eV)/k_BT_N]]^{-1}$. Here $T_N$ is the temperature in the N layer, and $-e$ is the electron charge.
The role of the FI layer is twofold: it acts as a spin filter with polarization\cite{moodera2007} $P=(G_\uparrow-G_\downarrow)/(G_\uparrow+G_\downarrow)$  and  causes the  spin-splitting of the  DoS in the S layer due to  the exchange coupling between the  localized magnetic moments of the FI and the conducting electrons of S \cite{deGennes,Tokuyasu1988,TMr}.  
These two features have been demonstrated in several experiments\cite{moodera1990,moodera2013,moodera2013prl,catelani2011,adams2013,beckmann2014}. Notice that, according to Eq.~(\ref{current_TE}), even in the absence of a voltage bias across the junction a finite current $I_{TE}$ can flow provided $T_N\neq T_S$, as demonstrated in Ref. \onlinecite{Ozaeta}.

\subsection{Electric and thermoelectric response of the TE}
\label{seccurrent}

Before analyzing  the role of a temperature bias across the TE, we  determine the current-voltage characteristics (IVCs) and differential conductance $G=dI_{TE}/dV$  of the NFIS junction. We set  a low temperature, $T_N=T_S=0.01T_c$, where $T_c=\Delta_0/(1.764k_B)$ is the critical temperature of the superconductor. 

The results obtained  from Eq.~(\ref{current_TE}) are summarized in Fig.~\ref{fig2}. 
Panels (a) and (b)   show the IVC and $G$, respectively, for a polarization of the barrier $P=50\%$ and different values of the spin-splitting exchange field $h_{exc}$. In panel (a) one clearly sees the deviation of the IVCs from those of a metal-insulator-superconductor (NIS) junction. For finite values of $h_{exc}$ there is a  sizeable subgap current [see Fig.~\ref{fig2}(a)] as a consequence of  
the spin-splitting of the DoSs in the S electrode. This splitting  manifests  itself also in the differential conductance $G$  [see Fig.~\ref{fig2}(b)], where the coherent peaks, 
usually appearing at $V=\pm\Delta/e$,  are now split in four peaks appearing at $V=(\pm\Delta \pm h_{exc})/e$. The asymmetry in the height of the coherent peaks stems from 
the spin polarization $P$ of the FI barrier [see Figs.~\ref{fig2}(c,d)] where we set $h_{exc}=0.4\Delta_0$ and the curves are calculated for different  values of $P$.  
Therefore from IVCs
one can estimate both  the polarization  of the barrier and the spin-splitting induced in S\cite{moodera1990}.

\begin{figure}[t!]
\includegraphics[width=\columnwidth]{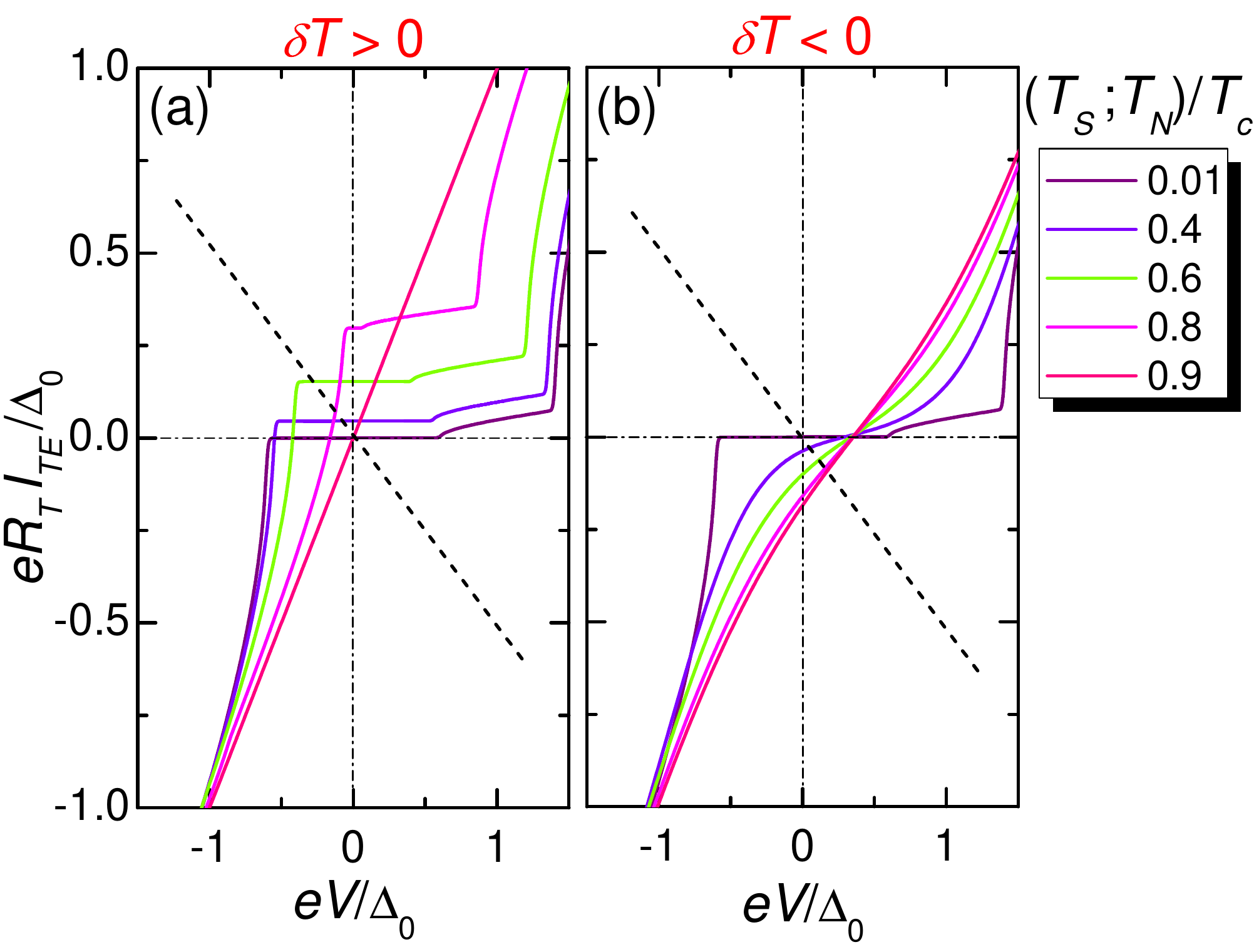}
\caption{(Color online) {\bf Characterization of the TE for $\mathbf{T_S\neq T_N}$} (a) $I_{TE}-V$ characteristic of TE calculated for several values of $T_S$ (see top legend) at $T_N=0.01T_c$, $P=0.9$, and $h_{exc}=0.4\Delta_0$. 
(b) The same as in panel (a) calculated for several $T_N$ values at $T_S=0.01T_c$. 
Dashed lines in panels (a) and (b) represent the current ($I_{JJ}$) flowing through the Josephson element when it is operated in the resistive regime, $I_{JJ}=-V/R_{JJ}$. 
\label{fig3} 
}
\end{figure}

We now assume a finite temperature difference between the electrodes\cite{note_th}, $\delta T=T_S-T_N$, and re-calculate the IVCs from Eq.~(\ref{current_TE}) for $h_{exc}=0.4\Delta_0$ and $P=0.9$.  
The results are shown in Figs.~\ref{fig3}(a) and \ref{fig3}(b), where we  keep  one of the electrodes at  temperature $0.01 T_c$  and change the other electrode temperature.   
The  curves in Fig.~\ref{fig3}  reveal two main properties of the IVC. 
First, the IVC  strongly depends on the amplitude of the temperature difference $\delta T$:  the larger the temperature difference, the larger is the current flow at low voltages. 
In the case that the S electrode is heated [see Fig.~\ref{fig3}(a)], this trend is limited by the reduced critical temperature $T_c^{\ast}<T_c$ of the superconductor originating from the presence of a  finite $h_{exc}$ which suppresses the $\Delta(T_S,h_{exc})$ calculated self-consistently.
When $T_S\rightarrow T_c^{\ast}$,  the TE is driven into the normal state with an ohmic characteristic [the red curve in  Fig.~\ref{fig3}(a)].  

Second, there is another interesting feature of the IVCs: they strongly depend on the \emph{sign} of $\delta T$. For the same value of  $|\delta T |$, the current  at  $V=0$ is  larger when the N electrode is colder than the S one, i.e., when $\delta T>0$.  In other words, the thermoelectric effect in the TE  strongly depends on the temperature difference. This feature was not investigated in previous works \cite{Machon,Ozaeta} in which only the linear response regime was discussed.  

\subsection{Measurement configurations}
\label{measurement}

 To use the TE element as a thermometer we need now to extract from the thermoelectrical signal  the temperature gradient present across  the TE junction. In order to do so we need to close the TE circuit over a generic load element modeled with a load resistance $R_L$ (see Fig.~\ref{fig1}).
In such a case the voltage $V$ developed across the TE for a given $\delta T$  is the solution of    
the  following  non-linear integral equation
 \begin{equation}
 I_{TE}(V,T_S,T_N,h_{exc},P)+\frac{V}{R_L}=0,
\label{totalcurrent}
 \end{equation}
where $I_{TE}$ is defined in Eq.~(\ref{current_TE}). 
The solution to the above equation is given by the point in which the dashed line (with slope proportional to $1/R_L$) in Figs.~\ref{fig3}(a) and \ref{fig3}(b) intersects the IVCs. 

For a Seebeck-like measurement one needs to maximize the thermovoltage opening the circuit, i.e., $R_L\to\infty$ and $I_{TE}=0$. 
For a Peltier-like measurement, one needs to maximize the current closing the circuit with a superconducting loop, i.e., $R_L=0$ and consequently $V=0$. In the case of   temperature-to-frequency conversion that we will discuss later,  one needs to include a Josephson element that operates in the dissipative regime with a load resistance $R_L=R_{JJ}$, which is  the total shunting resistance of the Josephson element. 

Independently of the chosen configuration, we assume to connect the TE to the detector with two superconducting arms S$_1$. 
In particular, we assume to place a tunnel barrier between S and S$_1$ to isolate the S element thereby ensuring 
 its description as a thermally homogeneous superconductor with a spin-split DoSs. We neglect here any influence of S$_1$ arms such that the current through the TE is described by Eq.~(\ref{current_TE}).
Superconductors S and S$_1$ are  Josephson coupled through the barrier so that no additional voltage drop will occur. 
Furthermore, we also assume the NS$_1$ junction to be a clean metallic contact, thereby contributing negligibly to the total resistance of the system and, for simplicity, we disregard the proximity effect induced into the N layer by the nearby contacted superconductor S$_1$ \cite{tinkham}.

\subsection{Linear response regime}
\label{linear}
In the linear response regime,  i.e., when the voltage $V$ and  temperature difference $\delta T\ll T\equiv (T_S+T_N)/2$ 
across the NFIS junction are small, Eq.~(\ref{totalcurrent})  reads~\cite{Machon,Ozaeta}.
\begin{equation}
\label{eq:linearI}
I_{TE}\approx I_{TE}^{lin}=\sigma V+ P\alpha\frac{\delta T}{T}\ ,
\end{equation} 
where 
\begin{equation}
\label{eq:sigma}
\sigma=\frac{1}{R_T}	\int_{-\infty}^{\infty} dE\frac{N_+}{4k_BT\cosh^2\left(\frac{E}{2k_BT}\right)}
\end{equation}
is the electric conductance, and 
 $\alpha$ is the thermoelectric Seebeck coefficient \cite{Ozaeta} defined as
\begin{equation}
\alpha
=\frac{1}{eR_T}\int_{-\infty}^{\infty} dE\frac{EN_-}{4k_BT\cosh^2\left(\frac{E}{2k_BT}\right)}.
\end{equation} 
which, in the linear regime, is connected to the Peltier coefficient $\Pi=\alpha/T$ by Onsager symmetry. 
Substituting Eq.~(\ref{eq:linearI}) in Eq.~(\ref{totalcurrent}) and solving respect to the thermovoltage across TE element 
one finds  
\begin{equation}
V^{lin}\simeq -P\alpha\frac{ R_{L}}{R_{L} \sigma+1}\frac{\delta T}{T}\; ,
\label{vlin}
\end{equation}
that is valid in the linear response regime  assuming a generic load resistance $R_L$.
We see immediately that the thermovoltage directly \emph{measures} the temperature gradient in TE.
Furthermore, for fixed load resistance, the achievable thermo-voltage $V$ increases with the polarization $P$. 
In an  open-circuit configuration ($R_{L}\rightarrow \infty$ and $I_{TE}=0$)  the TE thermo-voltage is maximal   being
\begin{equation}
\label{eq:Vlin}
V^{lin}\simeq -\frac{P\alpha}{\sigma} \frac{\delta T}{T}\ .
\end{equation}
For the closed-circuit ($R_L=0$ and $V=0$) instead the thermo-current is maximal being 
\begin{equation}
I_{TE}^{lin}\approx 
P\alpha \frac{\delta T}{T}\ . 
\end{equation}
Obviously we see that in the linear regime the open-circuit thermovoltage $V^{lin}$ is directly related to the closed-circuit thermocurrent, $I^{lin}=\sigma V^{lin}$. In particular, the dependence of the conversion efficiency on the polarization $P$ and the temperature gradient are  the same.
This simple picture drastically changes if one goes beyond the linear response regime, i.e., $\delta T \sim T$. We will see below that the  non-linear regime  is  essential in order to optimize the sensitivity for thermometry applications\cite{giazottormp2006}. 
%
 
\subsection{Intrinsic noise of TE element} 
\label{intrinsicnoise}

We now  address the zero frequency  noise performance of the NFIS junction. In this case, the main source of noise is the current noise ($S_I$) generated in the TE that is described by generalizing the expression derived in Ref.  \cite{golubev2001} in the presence of a ferromagnetic tunneling barrier:
\begin{equation}
\label{currentnoise_TE}
S_I=\frac{2}{R_T}\int_{-\infty}^{\infty}dE\left[N_++PN_-\right]\mathcal{M}(E,V,T_N,T_S),
\end{equation}
where 
\begin{equation}
\mathcal{M}=f_N(V,T_N)[1-f_S(T_S)]+f_S(T_S)[1-f_N(V,T_N)], 
\end{equation}
and the bias $V$ is given by the solution of Eq. (\ref{totalcurrent}). 
We note that the previous formula  describes  both thermal, i.e.,  $eV\ll T_N, T_S$, and shot noise, i.e., $T_N, T_S\ll eV$, and holds in the tunneling regime. 

Previous expression simplifies in the  linear response regime discussed before where we can neglect
any term $\mathcal{O}(\delta T)$ in Eq.~(\ref{currentnoise_TE}) finding the thermal noise 
\begin{equation}
S_I^{lin}=\frac{4}{R_T}\int_{-\infty}^{\infty}dE\left[N_++PN_-\right] f_N(T)[1-f_S(T)]\; ,
\end{equation}
which may be expressed as
\begin{equation}
\label{eq:linnoise}
S_I^{lin}=4 k_B T \sigma \; ,
\end{equation}
where $\sigma$ is the TE electric conductance of Eq.~(\ref{eq:sigma}). 
In the open-circuit configuration, it is  more convenient  to write the  voltage noise spectral density
\begin{equation}
\label{eq:linnoiseV}
S_V^{lin}=4 k_B T/\sigma\ .
\end{equation}
Below we show that in the non-linear regime  one can approximate  Eqs.~(\ref{eq:linnoise})-(\ref{eq:linnoiseV}) by substituting $\sigma$ by $1/R_d$, where $R_d$ is the  TE differential resistance.

\section{Temperature-to-voltage conversion}
\label{opencircuit}

\begin{figure}[t!]
\includegraphics[width=\columnwidth]{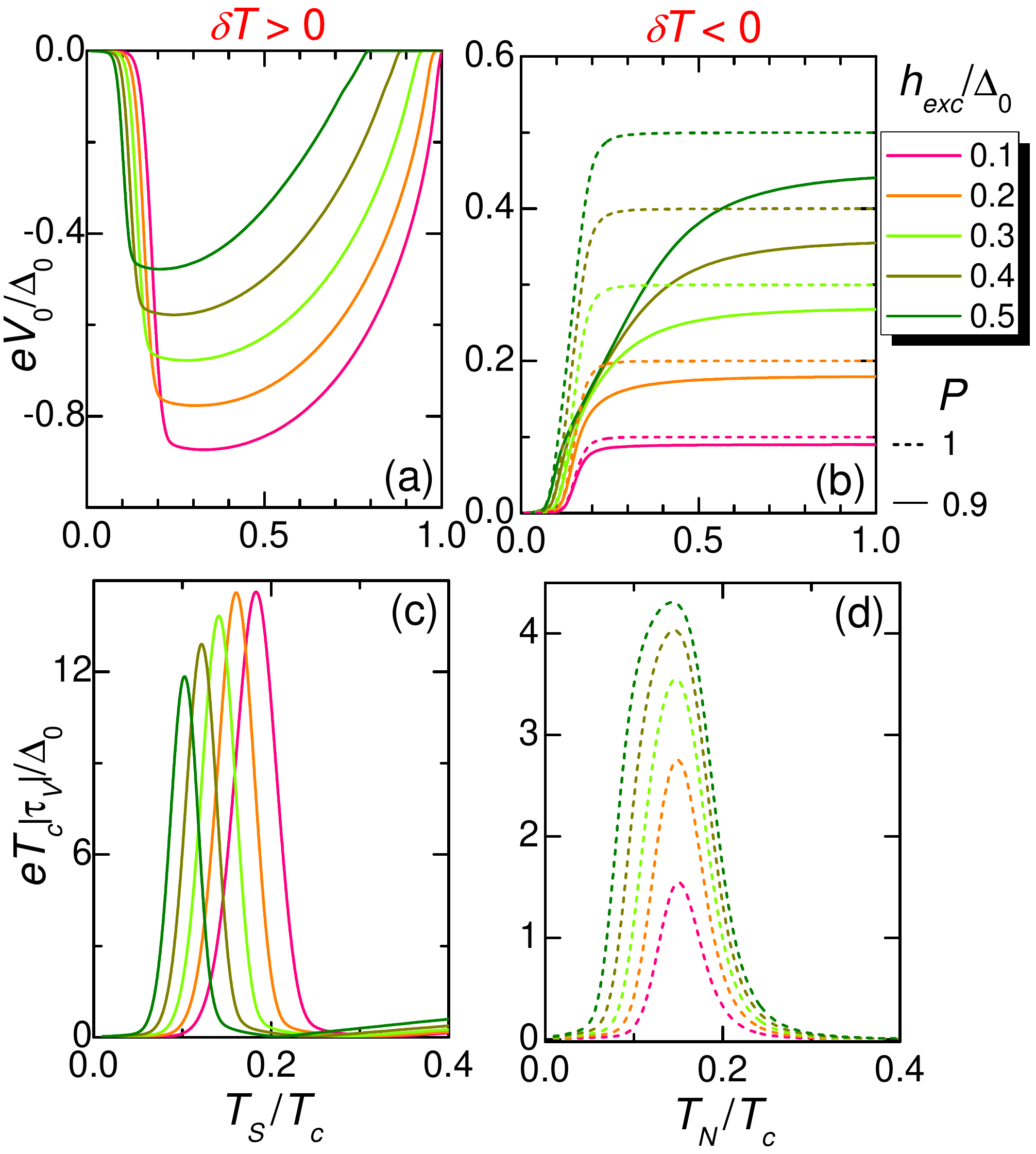}
\caption{(Color online) {\bf Open-circuit configuration.}
(a) Thermovoltage $V_0$ vs $T_S$ calculated for a few values of $h_{exc}$ (see legend) at $T_N=0.01T_c$ and $P=0.9$. (b) $V_0$ vs $T_N$ calculated for a few values of $h_{exc}$ at $T_S=0.01T_c$ and $P=0.9$ (full lines), or $P=1$ (dashed lines).
(c) and (d) show the absolute value of the corresponding  temperature-to-voltage transfer function $|\tau_V|$ calculated for the same values of $h_{exc}$.
\label{fig4}  
}
\end{figure}

When TE is in an \emph{open-circuit} configuration ($R_L\to\infty$), one can realize 
a \emph{temperature-to-voltage} conversion scheme. In such case no charge current flows  through the TE,
\begin{equation}
I_{TE}(V_0,T_S,T_N,h_{exc},P)=0, 
\label{opcirc}
\end{equation} 
then a voltage $V_0$ develops across the TE for  $\delta T\neq 0$. 
The value of $V_0$ can be obtained from the solution of Eq.~(\ref{opcirc}).  
The results are shown in the two upper panels of  Fig.~\ref{fig4}.  
Specifically, panel~\ref{fig4}(a) shows the dependence of $V_0$ on $T_S$ for different  values of $h_{exc}$,  $P=0.9$ and $T_N=0.01 T_c$.  
The increase of $T_S$, from the minimal temperature $T_N$, leads first to an enhancement of $|V_0|$. A further increase of $T_S$ leads to the suppression of the superconducting energy gap and a corresponding suppression of $V_0$. The voltage $V_0$ vanishes when when superconductivity is fully suppressed for $T_S\to T^*_c$.   
We note that $V_0$ reaches zero continuously owing  to the fact  that we have chosen  values of $h_{exc}$  for which the superconducting-normal state transition is of the second order\cite{buzdin}  

A different  temperature behavior of $V_0$ is obtained when S is kept at  $T_S=0.01T_c$ and  $T_N$ is varied,  as shown in  Fig.~\ref{fig4}(b).  
In particular, besides the obvious change of sign, $V_0$  grows monotonically  by increasing $T_N$  until it reaches an asymptotic value. 
It is important to stress that  the curves $V_0(T_N)$ depends strongly  on the polarization $P$  of the barrier [see Fig.~\ref{fig4}(b)]. 
In particular, the larger $P$ the larger is the thermovoltage $V_0(T_N)$ developed across the TE. By contrast, the $V_0(T_S)$ amplitude turns out to be almost unaffected by the value of $P$.

The different behaviors as a function of $\delta T$ allow one to reconstruct both the amplitude and direction of the thermal gradient in the TE element.
This further information could be eventually exploited to reconstruct the spatial position of a heating event, thereby opening interesting possibilities to build detector-like devices.

An useful figure of merit to estimate the performance of the TE is the temperature-to-voltage transfer function, $\tau_V=\partial V/\partial T$. The absolute value of this quantity is shown in Fig.~\ref{fig4}(c) for $\delta T>0$ and in Fig.~\ref{fig4}(d) for $\delta T<0$.  
We have normalized it to the natural unit $\Delta_0/eT_c$. 
In Fig.~\ref{fig4}(d) we show the case of barrier polarization selectivity $P=1$ which corresponds to the case with maximal possible transfer function at given $h_{exc}$. 
\begin{figure}[t!]
\includegraphics[width=\columnwidth]{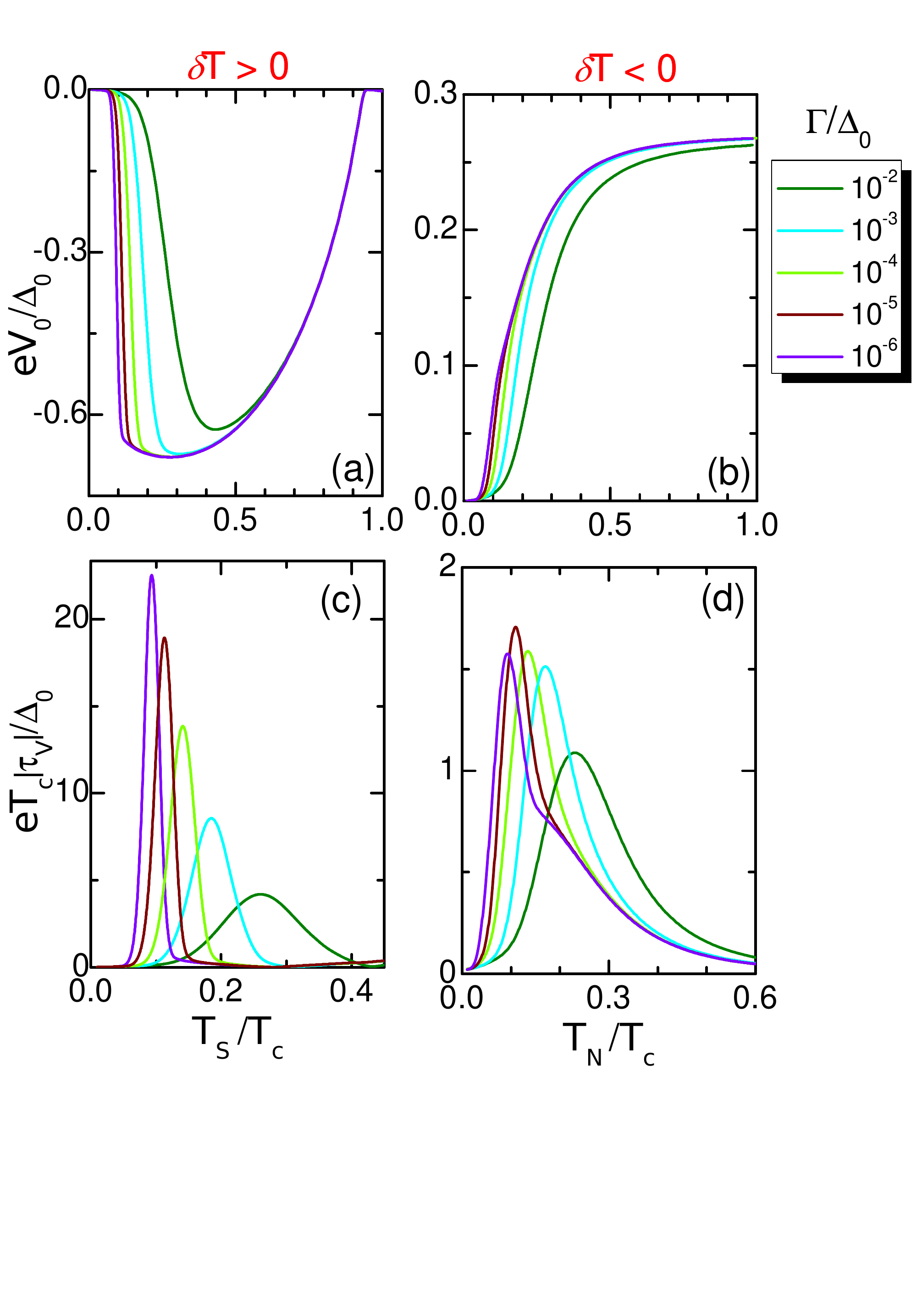}
\caption{(Color online) {\bf Open-circuit configuration.}
(a) Thermovoltage $V_0$ vs $T_S$ calculated for a few values of $\Gamma$ (see legend) at $T_N=0.01T_c$, $h_{exc}=0.3 \Delta_0$, and $P=0.9$. (b) $V_0$ vs $T_N$ calculated for a few values of $\Gamma$ at $T_S=0.01T_c$, $h_{exc}=0.3 \Delta_0$,  and $P=0.9$ (full lines). (c) and (d) show the absolute value of the corresponding  temperature-to-voltage transfer function $|\tau_V|$ calculated for the same values of $\Gamma$.
\label{figGamma}  
}
\end{figure}  

In order to show the impact of the broadening parameter we display in Fig. \ref{figGamma} the same quantities as in Fig. \ref{fig4} but calculated for fixed $h_{exc}=0.3\Delta_0$, $P=0.9$ and for different values of $\Gamma$ ranging from $10^{-6}\Delta_0$ up to $10^{-2}\Delta_0$ \cite{pekola2004,pekola2010,saira2012}. 
The overall qualitative behavior and the order of magnitude of the effect  is the same for  all these values.  
From a quantitative point of view, the temperature-to-voltage conversion turns out to be less effective the  larger the value of $\Gamma$. Throughout the paper  we assume 
$\Gamma=10^{-4}\Delta_0$ which is the typical value  for conventional Al-based superconducting junctions \cite{giazottormp2006,pekola2004}.

\subsection{Noise performance analysis for the \emph{open-circuit} configuration}
\label{sec:open_circuit_noise}

We now focus our analysis on  the noise performance of the temperature-to-voltage conversion with the NFIS junction. We need to convert it in a voltage noise assuming that load resistance $R_L\to\infty$. This means that the   voltage noise spectral density ($S_V$) generated across the TE, is\cite{Kogan}
\begin{equation}
\label{eq:S_V}
S_{V}(V_0,T_S,T_N,h_{exc},P)=S_I R_d^2,
\end{equation}
where the $R_d=\partial V_0/\partial I_{TE}$ is the differential resistance of the TE,  and the bias $V_0$ is given by the solution of Eq. (\ref{opcirc}). 

\begin{figure}[t!]
\includegraphics[width=\columnwidth]{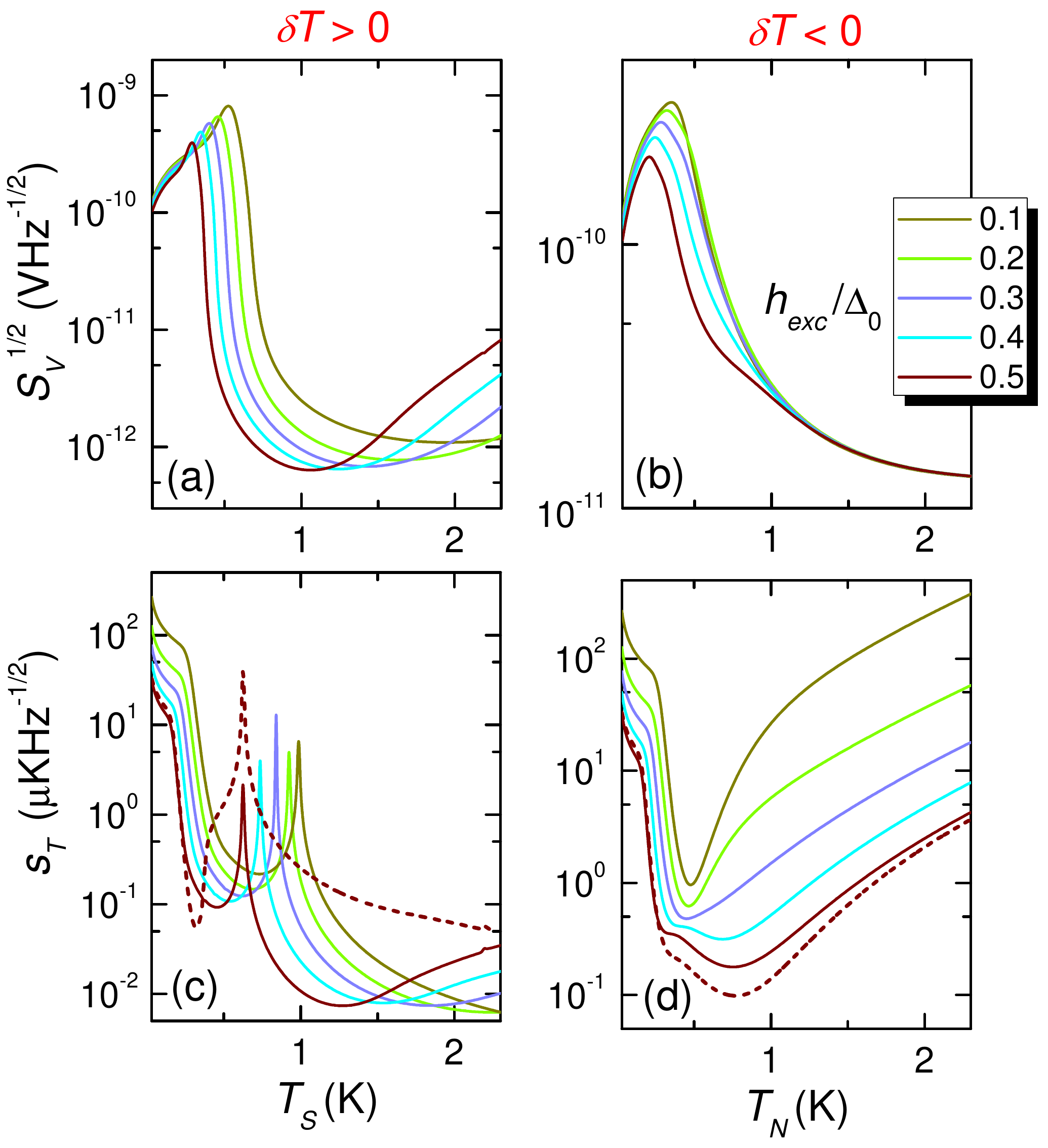}
\caption{(Color online) {\bf Open-circuit configuration.} (a) Square root of the voltage noise spectral density $S_V^{1/2}$ vs $T_S$ calculated at $T_N=0.01 T_c$ for a few values of $h_{exc}$ (see legend). 
(b) $S_V^{1/2}$
 vs $T_N$ calculated at $T_S=0.01 T_c$ for the same $h_{exc}$ values. (c) Temperature sensitivity $s_T$ vs $T_S$ calculated for the case of panel (a). (d) $s_T$ vs $T_N$ calculated for the case of panel (b). Dashed lines show $s_T$ as calculated using the approximate formula Eq.\ref{eq:appSv} for $h_{exc}/\Delta_0=0.5$. In all these calculations we set $P=0.98$, and $T_c=3$K.
}
\label{fig5}
\end{figure}     
In Fig.~\ref{fig5}(a)-(b) the square root of noise spectral density ($\sqrt{S_V}$) is displayed for a TE element with a barrier characterized by a realistic value of polarization $P=0.98$. This spin-filter efficiency is representative for EuO or EuS FI barriers, \cite{moodera2008} and we assume a superconductor with $T_c=3$K which would be implementable with ultra-thin Al films \cite{moodera2013,moodera2013prl,catelani2011,adams2013}.

We find that for $\delta T>0$ the minimal noise value is obtained in the non-linear regime $T_N\ll T_S \lesssim T_c$ where the voltage noise can be as low as  $\sim 600$fV Hz$^{-1/2}$, and is two orders of magnitude lower than the equilibrium case $\delta T/T\ll 1$, where $T=(T_S+T_N)/2$ is the average temperature. For $\delta T<0$, the noise performance is worse being at best a few tens of pV Hz$^{-1/2}$ for the non-linear regime $T_S\ll T_N \sim T_c$. 

The intrinsic temperature noise (temperature sensitivity) per unit bandwidth of the thermometer ($s_T$) is related to the voltage noise spectral density as
\begin{equation}
\label{eq:StV}
s_T=\frac{\sqrt{S_V}}{|\tau_V|}.
\end{equation}
In Fig.~\ref{fig5}(c)-(d) we show the temperature noise $s_T$ for the open-circuit configuration for the two cases of Fig.~\ref{fig5}(a)-(b). The differences between the voltage spectral density is entirely given by the transfer function which is highly non-linear as a function of $\delta T$. We notice that in the linear regime $|\delta T|/T\ll 1$ the temperature noise, given by Eq.~(\ref{eq:linnoiseV}), is  a few tens of $\mu$K Hz$^{-1/2}$. 
The maximum temperature sensibility is obtained in the nonlinear regime, i.e., $|\delta T|/T \gg 1$ where the temperature noise can be as low as $8$nK Hz$^{-1/2}$ coinciding with the minimal voltage noise. By contrast, for the case $\delta T<0$, the best noise performance is around $180$nK Hz$^{-1/2}$. 

It is interesting to observe the scaling behaviour of noise power as function of the junction normal-state resistance 
$R_T$. Indeed as $S_I\propto 1/R_T$ [see Eq.~(\ref{currentnoise_TE})], from Eq.~(\ref{eq:S_V}) one can conclude that $S_V\propto R_T$ since $R_d\propto R_T$. At the same time there is no scaling behaviour of the transfer function, $\tau_V=\partial V_0/\partial T$. 
This may be easily inferred, for instance, from the expression of the thermovoltage $V_0$ in the linear regime, Eq.~(\ref{eq:Vlin}), since $V^{lin}\propto \alpha/\sigma$ which is the ratio of two quantities with the same scaling $1/R_T$. 
Alternatively one can deduce it from the relation between open-circuit voltage an the temperature difference, $\delta T$ Eq.~(\ref{opcirc}), where $R_T$ enters only as an overall prefactor. We conclude that  $s_T\propto \sqrt{R_T}$ which shows immediately that the reduction of TE  resistance would be beneficial for increasing the sensitivity in temperature measurement. 

These considerations suggest that in the non-linear regime  the differential resistance $R_d$ takes the role of $R_T$.
Indeed one can guess a way to generalize  Eq.~(\ref{eq:linnoiseV}) to the non-linear regime 
by replacing the linear conductance $\sigma$ by the differential conductance $1/R_d$ such that
\begin{equation}
\label{eq:appSv}
S_V\approx 4 k_B T R_d 
\end{equation}
where the temperature is taken as the average $T=(T_S+T_N)/2$.   The previous expression should converge to the linear result when $|\delta T|/T\ll 1$. 
The full numerical results of  Fig.~\ref{fig5}(c)-(d) demonstrate the accuracy  of Eq. (\ref {eq:appSv}) shown as  dashed lines. 
This shows that the noise performance is essentially characterized by the dependence of differential resistance $R_d$. 
Therefore this approximation is extremely useful to estimate the noise performance with the knowledge of the differential resistance $R_d$ only.

It is important to emphasize that in a realistic measurement scheme the temperature sensitivity of the device is  limited by the amplifying chain. 
Indeed, in general, the voltage signal must be amplified with a low-noise preamplifier which is characterized by its intrinsic voltage noise. 
The preamplifier noise may degrade the total noise performance. In particular, assuming for the pre-amp a square root spectral density of $\sim1$nV Hz$^{-1/2}$, it is clear that it will dominate over the intrinsic voltage noise of the signal which can be smaller by a few orders of magnitude [see Fig.~\ref{fig5}(a)-(b)]. 
The preamplifier is therefore the main bottleneck to the temperature detection in this configuration scheme although it has the advantage of suppressing the noise non-linearities over the considered temperature window. The realistic performance of this measurement scheme will be roughly $\sim10\mu$K Hz$^{-1/2}$. 
This limitation can be overcome by exploiting a close-circuit configuration, as  discussed in the next section.

\section{Temperature-to-current conversion}
\label{closedcircuit}
\begin{figure}[t!]
\includegraphics[width=\columnwidth]{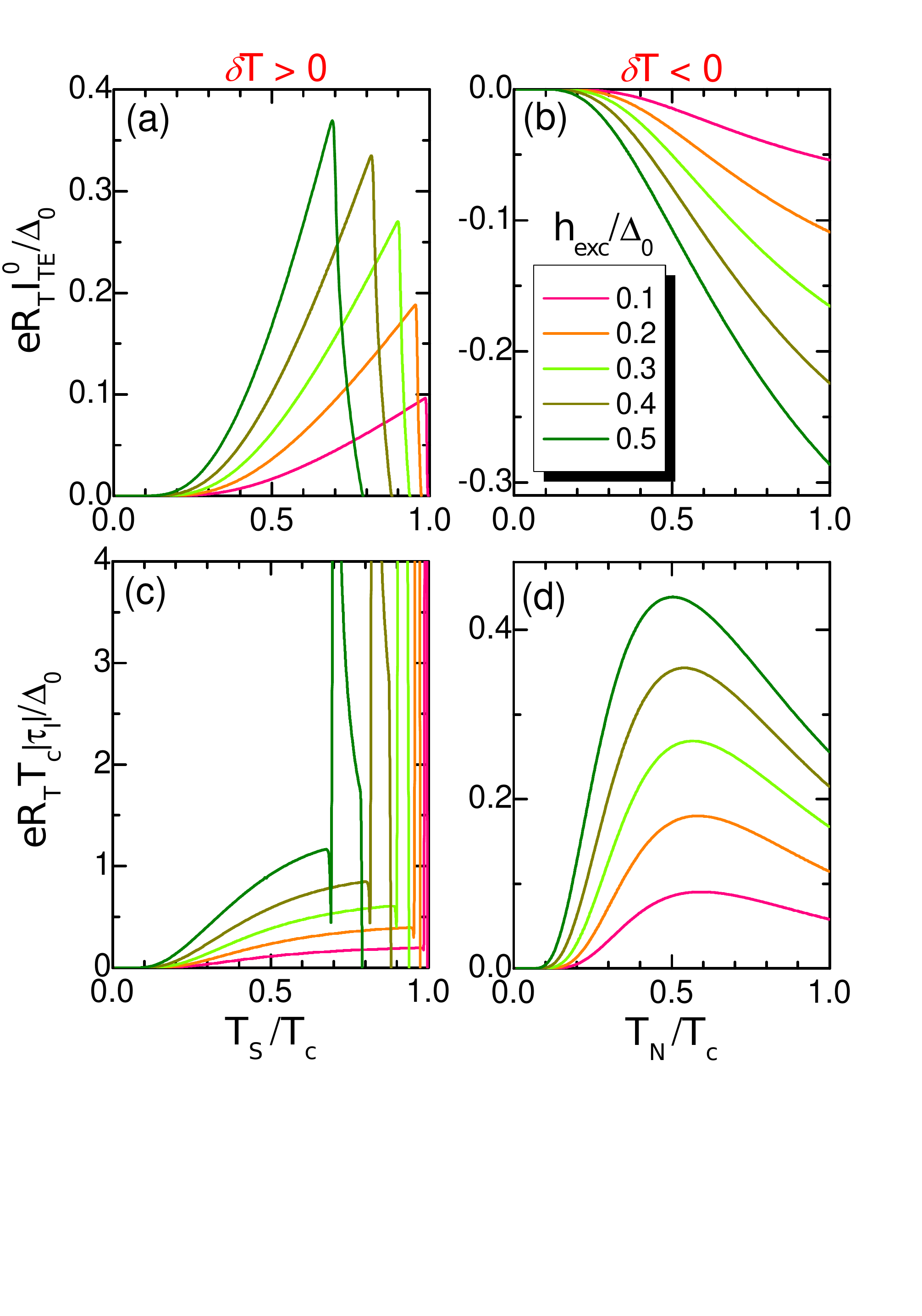}
\caption{(Color online) {\bf Closed-circuit configuration} (a)  Thermocurrent $I_{TE}^0$ vs $T_S$ calculated for a few values of $h_{exc}$ at $T_N=0.01 T_c$. 
(b) $I_{TE}^0$  vs $T_N$ calculated for the same values of $h_{exc}$.
(c) and (d) show the absolute value of the temperature-to-current transfer function $|\tau_I|$ vs temperature for the panel (a) and (b), respectively, calculated for the same values of $h_{exc}$.  In all these calculations we set $P=0.98$. 
}
\label{fig6}
\end{figure}   

Hereafter we analyse the performance of a  closed-circuit configuration which correspond to \emph{temperature-to-current} conversion. In this setup the TE current $I_{TE} (V=0)= I_{TE}^0$ is given by Eq.~(\ref{current_TE}) which depends only on $T_S$ and $T_N$.

In Fig.~\ref{fig6}(a) we show how the current depends on $T_S$ for different values of $h_{exc}$ keeping fixed the barrier polarization $P=0.98$ and  $T_N=0.01T_c$. The general behaviour has "shark-fin" shape which increases in amplitude with the $h_{exc}$. After reaching a maximum at $T_S^*$ the $I_{TE}^0$ decreases with $T_S$ until the critical temperature $T_c^*$ is reached and the superconductivity is completely suppressed. 

If we fix $T_S=0.01T_c$, by changing $T_N$ we get for $\delta T<0$ an obvious opposite sign for the thermocurrent $I_{TE}^0$, and an absolute value of the thermocurrent $|I_{TE}^0|$ which monotonously increases by enhancing $|\delta T|$. It finally saturates to the maximal value
\begin{equation}
I_{TE,max}^0=\frac{1}{2eR_T}\int_{-\infty}^{\infty} dE\left[N_++PN_-\right]\mathrm{sgn}(E),\;
\end{equation} 
which is easily obtained from the general expression of the TE current, Eq.~(\ref{current_TE}), by taking the limit $T_S\to0$ and $T_N\to\infty$ with $V=0$. 
From this, we can conclude that  an arbitrary enhancement of $|\delta T|$ is not of particular benefit to increase the current signal.

In Fig.~\ref{fig6}(c)-(d) we show the absolute value of the temperature-to-current transfer function $\tau_I=\partial I_{TE}^0/\partial T$ respectively for the case (a) and (b) of the same figure. For $\delta T>0$, the transfer function has two different behaviors depending if $T_S$ is smaller or larger than $T_S^*$. On the other hand one sees that,  independently of the sign of $\delta T$,   the  transfer function is maximized  in the non-linear regime $|\delta T|\sim T$. This is an important issue in order to increase the temperature sensitivity. 
\begin{figure}[t!]
\includegraphics[width=\columnwidth]{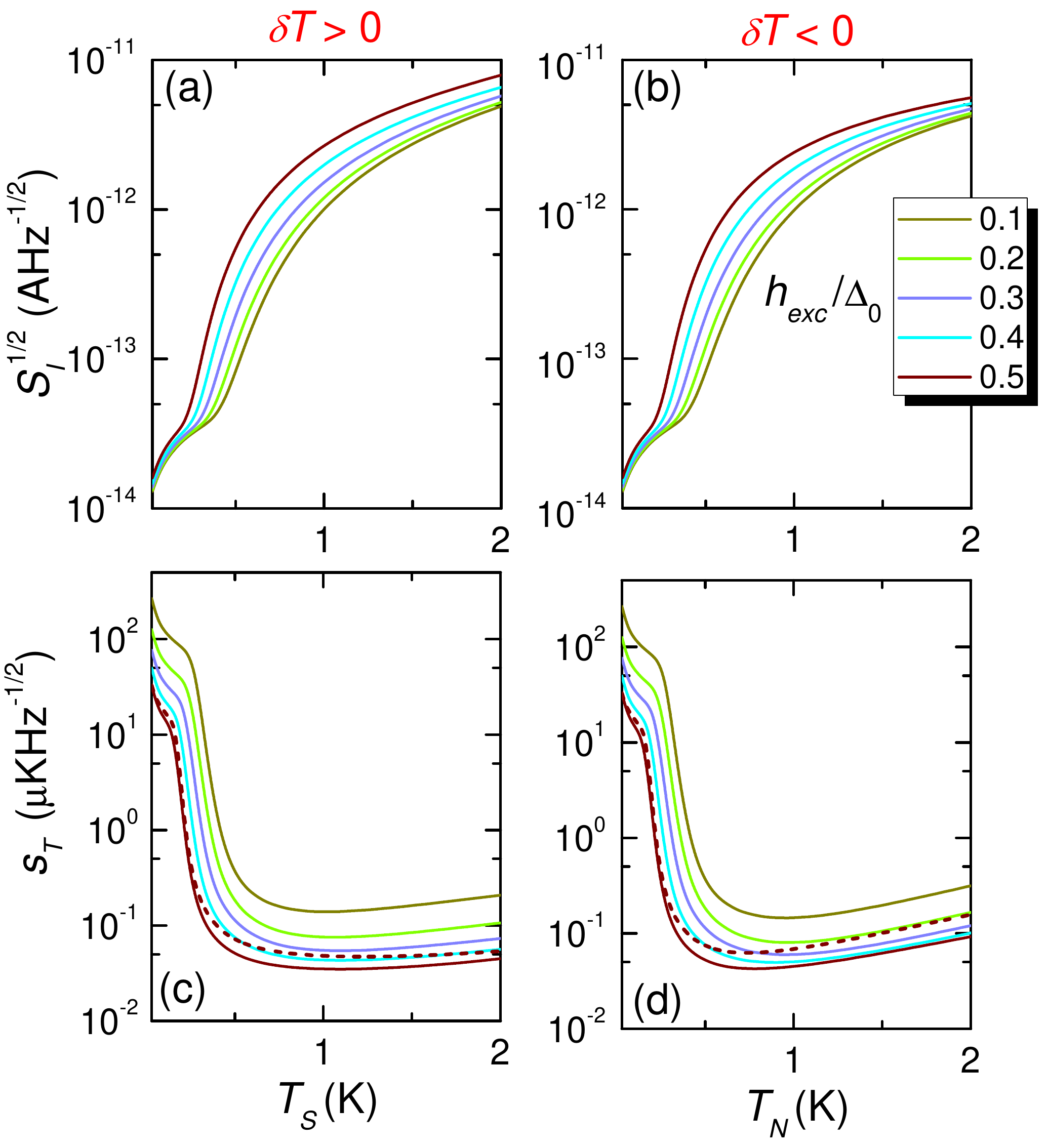}
\caption{(Color online) {\bf Closed-circuit configuration}  (a) Square root of the current noise spectral density $S_I^{1/2}$ vs $T_S$ for $T_N=0.01T_c$ and for different values of $h_{exc}$ (see legend). (b) $S_I^{1/2}$ vs $T_N$ for $T_S=0.01T_c$ for the same values of $h_{exc}$. (c) $s_T$ vs $T_S$ for the case of panel (a). (d) $s_T$ vs $T_N$ for the case of panel (b). Dashed lines shows $s_T$ as calculated using the approximate formula Eq. (\ref{eq:appSI})  for $h_{exc}/\Delta_0=0.5$.  In all these calculations we set $P=0.98$ and $T_c=3$K.}
\label{fig7}
\end{figure}

\subsection{Noise performance analysis for the \emph{closed-circuit} configuration}

In Fig.~\ref{fig7}(a)-(b) the current  noise $S_I$ of the closed-circuit configuration is shown as obtained from Eq.~(\ref{currentnoise_TE}) with $V=0$. 
The current noise as a function of 
$\delta T$ is minimized  in the linear regime obtaining $\sim15$fAHz$^{-1/2}$, and grows by increasing $|\delta T|$. 
The noise behaviour of the closed-circuit configuration it less affected by the sign of $\delta T$ in comparison to the open-circuit one (see Sec.  \ref{sec:open_circuit_noise}). The current noise increases with $h_{exc}$ since also the average current $I_{TE}^0$ increases in such a case [see Fig.~\ref{fig6}(a) and (b)].   

The intrinsic temperature noise per unit bandwidth of the thermometer $s_T$ in this configuration is given by
\begin{equation}
\label{eq:StV2}
s_T=\frac{\sqrt{S_I}}{|\tau_I|},
\end{equation} 
where $|\tau_I|$ is the temperature-to-current transfer function discussed before. 
The same scaling behaviour shown before for $s_T\propto \sqrt{R_T}$  still holds in this configuration since now $S_I\propto 1/R_T$ but $\tau_I\propto 1/R_T$. 
Consequently also for this case the minimization of $R_T$ would be, in general, beneficial for improving noise performance. 
 As in the previous section one can try to generalise this argument for the non-linear regime by replacing $R_T$ with $R_d$. 
Since $R_d$ is largely reduced in comparison to the linear regime value $1/\sigma$, one can expect an increase of the  
noise  in the non-linear regime. At a first glance, this does not look plausible since  the current noise is in general higher [see Fig.~\ref{fig7}(a)-(b)]. However, our guess seems to be  correct, as shown  in Fig.\ref{fig7}(c) and (d), where the  temperature noise is minimized for large values of $\delta T$.

The lowest intrinsic noise $\sim 35$nK Hz$^{-1/2}$ is obtained in the non-linear regime for $\delta T\approx 1K$, when $P=0.98$ and $T_c=3$K are chosen.
The main difference with the open-circuit configuration (see Fig. \ref{fig5}) is that in the present situation  the noise depends weakly on the sign of $\delta T$ in a wide temperature region. Moreover, in contrast to the open-circuit configuration, the noise shows a rather smooth behavior. 

As pointed out above, the behaviour of the current noise in the non-linear regime can be  approximated  by the expression 
\begin{equation}
\label{eq:appSI}
S_I\approx \frac{4 k_B T}{R_d},
\end{equation}
where the linear conductance $\sigma$ of Eq. (\ref{eq:linnoise}) is replaced by the inverse differential resistance, $1/R_d$. 
In Fig. \ref{fig7}(c) and (d) we show  (dashed lines)  this approximation  for the case where we expect the largest nonlinearities, i.e., for $h_{exc}/\Delta_0=0.5$.  
We can thus conclude that this simple formula gives a fairly accurate description of $S_I$ in the non-linear regime.  

In terms of overall temperature noise  the closed-circuit configuration has two advantages:
First, the smooth behaviour of temperature noise makes it more attractive than the  open-circuit configuration.
Second, while the ideal noise is better for the open-circuit configuration (see Figs. \ref{fig5} and  \ref{fig7}), one needs to evaluate the  total noise of the measurement which includes  the addition of the  preamplifier noise. 
The latter, as discussed in the previous section,  strongly degrades the resulting noise figure. The close-circuit configuration offers a way to overcome this limitation, as discussed in the following. 

Specifically, we propose  to measure the current signal by coupling  the closed circuit via  a mutual inductance $M$ to a SQUID. The latter   measures the flux  generated by the  current circulating in the thermoelectric circuit. 
The total temperature sensitivity,  which includes now the SQUID noise, can be written as
\begin{equation}
\label{eq:Stphi}
s_T^{\rm tot}=\frac{\sqrt{S^{TOT}_\phi}}{|\tau_\phi|}=\frac{\sqrt{S_I+(S^{SQUID}_\phi/M^2)}}{|\tau_I|},
\end{equation} 
where the temperature-to-flux transfer function is $\tau_\phi=M\tau_I$, and the TE flux spectral density 
$S_\phi^{TE}=M^2 S_I$ is added to the SQUID noise ($S^{SQUID}_\phi$) to give the total flux noise, $S^{TOT}_\phi=S_\phi^{TE}+S^{SQUID}_\phi$. 
The square root of the flux noise for high-quality commercial SQUID can be as low as  $\sqrt{S^{SQUID}_\phi}\sim10^{-7}\Phi_0$ Hz$^{-1/2}$, which is then converted into an effective circulating current noise in the thermoelectric circuit of $\sim20$ fA Hz$^{-1/2}$ by dividing it with a typical value for the mutual inductance $M=10^{-8}$H. 
By looking at Fig.~\ref{fig7}(a) and (b) we immediately see that the intrinsic TE current noise will, in general, dominate over the SQUID noise almost everywhere in the non-linear regime where we can achieve the best sensitivity. 
Therefore, the temperature noise of this measurement scheme is only limited by the intrinsic TE noise mechanisms, and can be as low as $\sim 35$nK Hz$^{-1/2}$ for a moderate temperature non-linearity (see Fig.\ref{fig7}).

\section{Temperature-to-frequency conversion}
\label{temptofreq}

\begin{figure}[t!]
\includegraphics[width=\columnwidth]{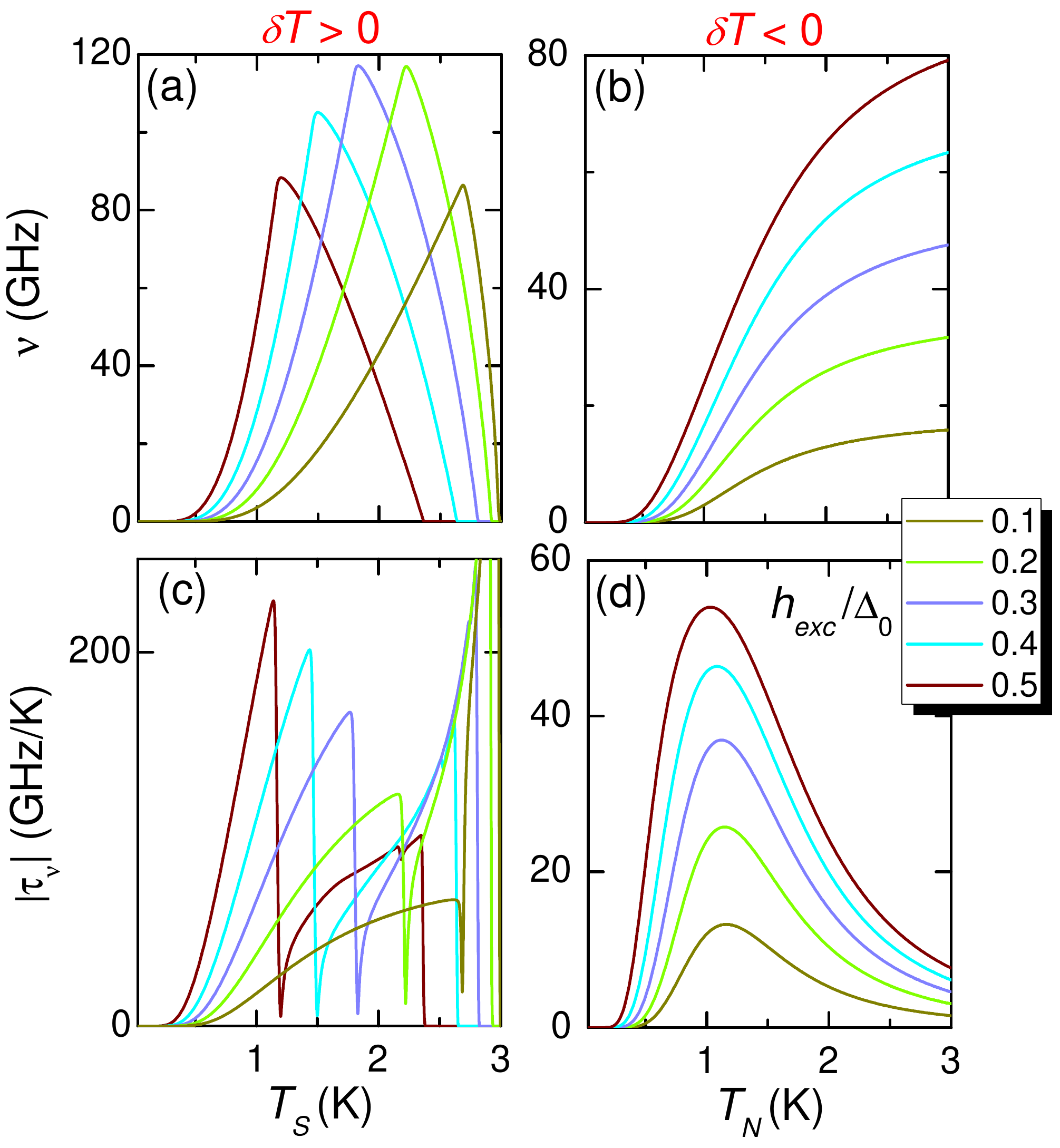}
\caption{(Color online) {\bf Characterization of the temperature-to-frequency converter} (a) Frequency $\nu$ vs $T_S$ calculated for a few values of $h_{exc}$ at $T_N=0.01 T_c$. (b) $\nu$ vs $T_N$ at $T_S=0.01T_c$ calculated for the same $h_{exc}$ values as in panel (a). (c) and (d) show the absolute value 
transfer function $|\tau_{\nu}|$ of panel (a) and (b), respectively, calculated for the same values of $h_{exc}$. In all these calculations we set $P=0.98$, $R_T/R_{JJ}=0.2$, and $T_c=3$K.
\label{fig8} 
}
\end{figure}
We now focus on the \emph{temperature-to-frequency} conversion process. This conversion is achieved with the device  
sketched  in Fig.~\ref{fig1} where the thermovoltage generated across the TE is applied to a generic Josephson element. 
The latter is set to operate in the dissipative regime when $I_{JJ}=V/R_{JJ}$, where $R_{JJ}$ is the total shunting resistance of the Josephson element.
In this case, there is a time-oscillating current through the Josephson element with a frequency equal 
to the Josephson frequency, $\nu=|V|/\Phi_0$. As discussed above, the value of $V$ depends 
on the temperature difference $\delta T$ across  the TE,  and therefore the frequency emitted by the Josephson junction is  a measure of $\delta T$.    

In order to quantify the temperature-to-frequency conversion effect, one has to determine the voltage  $V$ developed for any given $\delta T$ imposed across the TE which satisfies  Eq.~(\ref{totalcurrent}) with a finite load resistance $R_L=R_{JJ}$. 
This configuration is intermediate between the open-circuit and the closed-circuit setup discussed above.
In the following, we consider the  case where $R_T/R_{JJ}=0.2$ in order to produce a  detectable frequency signal between $10$ GHz and fractions of THz. 

The frequency-to-temperature performance of this configuration is shown in Fig.~\ref{fig8}. We again used the spin-filter efficiency $P=0.98$ and the critical temperature $T_c=3$K adopted in the previous sections. 
Panels~\ref{fig8}(a) and \ref{fig8}(b) show the frequency generated by the Josephson element, for positive and negative $\delta T$, respectively. 
In the linear response regime, the TE thermovoltage depends only on $|\delta T|$ as can be seen from Fig. \ref{fig8}.
The information about the sign of $\delta T$ is eventually recovered only for the non-linear regime.

If $T_N$ is kept at $0.01T_c$ the maximum frequency is achieved around  $h_{exc}\approx 0.2 \Delta_0$ for  $T_S\approx 0.75 T_c$, and obtains values as large as $\sim 120$GHz. 
If $T_S$ is kept at low temperature, $\nu$ increases monotonically by increasing  both  $T_N$ and/or $h_{exc}$ and obtains a maximum of $\sim 80$GHz

In the present setup an important  figure of merit of the structure is represented by the temperature-to-frequency transfer function, $\tau_\nu=\partial \nu/\partial T$, plotted in absolute value in Figs.~\ref{fig8}(c) and \ref{fig8}(d).
In particular, $|\tau_\nu|$ exceeding $200\,$GHz/K around $T_S\sim 1$K  can be achieved for $h_{exc}=0.5\Delta_0$ by heating S, while $|\tau_\nu|$ up to $\sim 55\,$GHz/K can be achieved with the same values by heating  N.      

\begin{figure}[t!]
\includegraphics[width=\columnwidth]{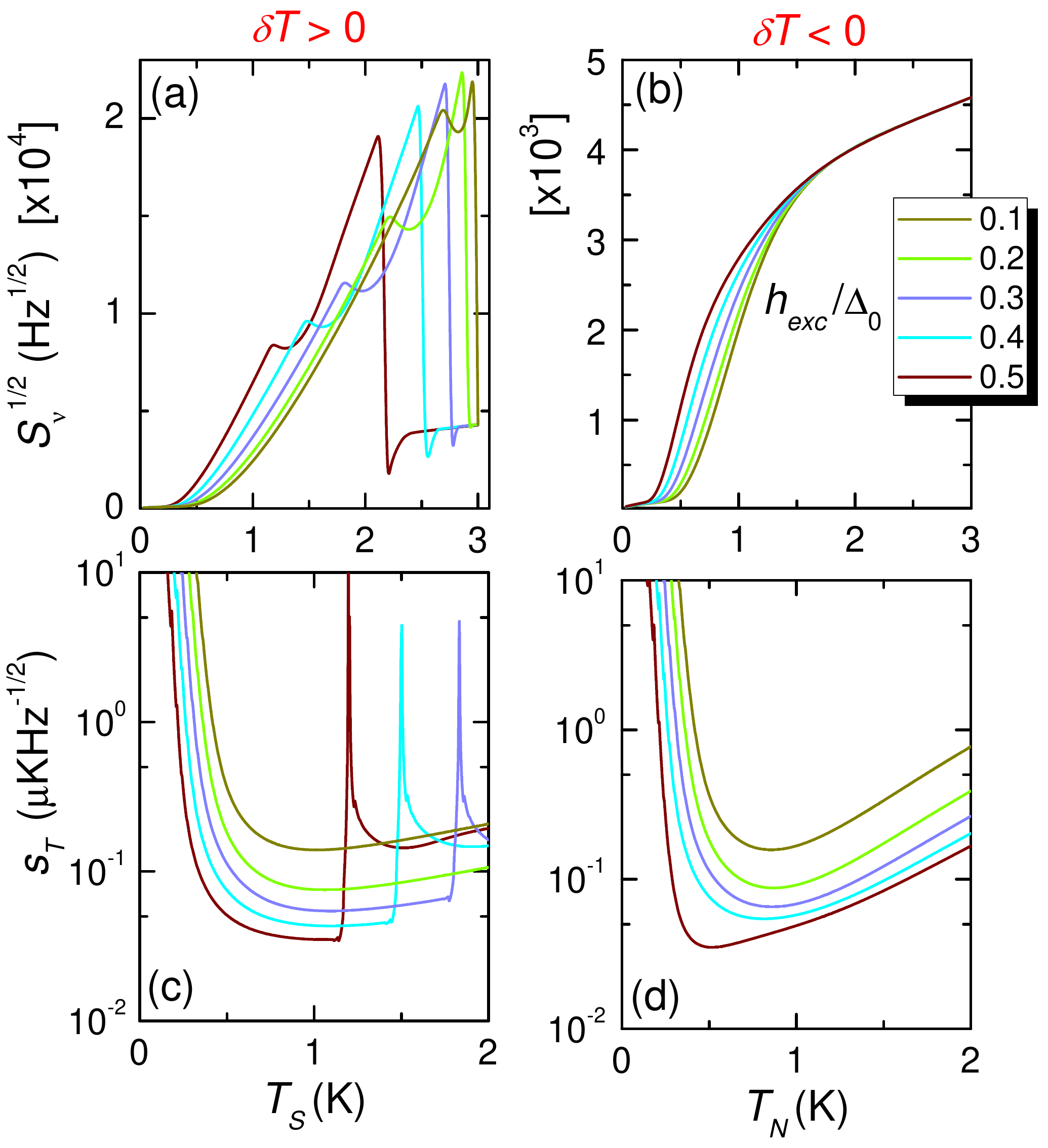}
\caption{(Color online) {\bf Noise performance of the temperature-to-frequency converter} (a) Square root of the frequency noise spectral density $S_{\nu}^{1/2}$ vs $T_S$ calculated at $T_N=0.01 T_c$ for a few values of $h_{exc}$. (b) $S_{\nu}^{1/2}$
 vs $T_N$ calculated at $T_S=0.01 T_c$ for the same $h_{exc}$ values. (c) Temperature sensitivity $s_T$ vs $T_S$ calculated at $T_N=0.01 T_c$ for a few values of $h_{exc}$. (d) $s_T$ vs $T_N$ calculated at $T_S=0.01 T_c$ for the same $h_{exc}$ values. In all these calculations we set $P=0.98$, $T_c=3$K, $R_T/R_{JJ}=0.2$, and $R_T=1\Omega$.
\label{fig9} 
}
\end{figure}

\subsection{Noise performance}
\label{noise}

In the temperature-to-frequency conversion process, the noise is determined by the bias fluctuations generated from the current noise via the load resistance seen by the TE, i.e., the parallel between the Josephson element total resistance $R_{JJ}$ and the TE resistance $R_d$:   $\mathcal{R}=R_d R_{JJ}/(R_d+R_{JJ})$.
Note that the differential resistance $R_d=\partial V_0/\partial I_{TE}$ is calculated from the solutions of Eq.~(\ref{totalcurrent}) where $R_L=R_{JJ}$.
The important quantity is represented by the frequency noise spectral density ($S_{\nu}$) which can be expressed as 
\begin{equation}
S_{\nu}=\frac{S_I\mathcal{R}^2}{\Phi_0^2}\ .
\end{equation}
Finally, the intrinsic temperature noise per unit bandwidth of the thermometer ($s_T$) is related to the frequency noise spectral density as
\begin{equation}
s_T=\frac{\sqrt{S_{\nu}}}{|\tau_\nu|}.
\end{equation}

Figure~\ref{fig9}(a) and (b) show the calculated square root of the frequency noise spectral density $S_{\nu}$ for positive and negative $\delta T$, respectively, calculated for the same parameters as in Fig.~\ref{fig8}, and for $R_T=1\Omega$. 
In particular, for positive $\delta T$,  the noise spectrum $S_{\nu}^{1/2}$ shows a non-monotonic behavior with a maximum at intermediate temperatures, and suppression at higher $\delta T$. By contrast, for $\delta T<0$  the noise spectrum grows monotonically with $|\delta T |$, and it is less influenced by $h_{exc}$.

The behavior of $s_T$ is displayed in Fig.~\ref{fig9}(c) and (d). 
At small  $|\delta T|$, in the linear regime, the noise sensitivity is given by several tens of $\mu$K Hz$^{-1/2}$. 
By increasing $|\delta T |$ the growth of $S_{\nu}^{1/2}$ [Figs. \ref{fig9}(a) and (b)] is advantageously compensated by the enhancement of $\tau_\nu$ [see Fig.~\ref{fig8}(c) and (d)]. 
The best noise performance is obtained when $|\tau_\nu |$ it is quite near its maximum.
The values of $s_T\sim 35\,$nK Hz$^{-1/2}$ is obtained around 1K for $h_{exc}=0.5\Delta_0$. After the minimum of $s_T$, for $\delta T>0$, we see a peak due to the divergence of $|\tau_\nu|^{-1}$, i.e., the vanishing of the transfer function, as shown in Fig.~\ref{fig8}(c). Differently, for $\delta T<0$ one observes a smooth increase of $s_T$ determined by the progressive reduction of the transfer function [see Fig.~\ref{fig8}(d)] which is consequence of the saturation of the frequency when $T_N\to\infty$. 
We conclude by noticing that, also for this case, the best noise performance is obtained in the non-linear regime for $|\delta T|\lesssim 1$K where $s_T$ is almost independent of the sign of $\delta T$. This is essentially due again to the fact that $R_d$ is strongly reduced in the non-linear regime.

In this configuration the power of the generated frequency signal  might be somewhat low. Anyway, one can deploy the standard techniques in order to increase the emission power by connecting in parallel arrays of JJs.\cite{wengler1994,barbara1999}. The only limiting factor in that case would be the power that the TE element could sustain and transfer to the JJs. A rough estimate shows that when $R_T/R_{JJ}=0.2$ and $R_T=1\,\Omega$ the TE could produce a power of the order of $\sim 100$pW...10nW which would be high enough to make the $10-100$GHz signal generated by the Josephson junction detectable.

\section{Summary}
\label{summary}

In summary, we have theoretically investigated a thermoelectric structure based on a normal metal-ferromagnetic insulator-superconductor (NFIS) junction. We fully characterize the thermoelectrical properties of the TE both in the linear and non-linear regimes. We assumed different measurement configuration as determined by the load resistance value.
In particular, we showed that by exploiting realistic materials such as EuS or EuO (providing polarization $P$ up to $\sim 98\%$) in combination with  superconducting Al thin films the device is able to provide remarkable temperature noise performances. 
We find that in the open circuit configuration, where the temperature signal is returned via the \emph{Seebeck thermo-voltage}, the lowest achievable intrinsic noise of $\sim 10$nKHz$^{-1/2}$ is limited by the amplifying chain.
On the other side, we found that in the closed-circuit configuration, where the temperature information is encoded in the \emph{Peltier thermo-current}, one can detect the signal via a low-noise flux measurement of an inductively-coupled SQUID. In such case the temperature noise performance is mainly determined by intrinsic noise mechanisms, with the best value of $\sim 35$nK Hz$^{-1/2}$ achievable with state-of-art SQUID technology. Interestingly,  we identified in the differential resistance $R_d$ of the TE one of the main factor that determines the intrinsic noise performance of the system. 
This explain why the best noise performances are obtained in the non-linear temperature regime since for that regime $R_d$ is strongly suppressed. This is a non-trivial consequence of the strong non-linearities peculiar of the NIFS junction. 

We finally discuss a \emph{temperature-to-frequency} converter where the obtained thermovoltage is converted through a dissipative Josephson junction into a high frequency signal in the frequency window spanning from a few GHz up to $\sim 10^{11}$Hz.
In particular, we have shown that the device allows for the generation of Josephson radiation at a frequency that depends on both the \emph{amplitude} and  \emph{sign} of the temperature difference across the NFIS junction therefore opening the route for high-frequency detection associated to high temperature sensitivity. 
Frequencies up to $\sim 120$GHz and large transfer functions (i.e., up to $200$GHz/K) around $\sim 1-2$K can be obtained in a structure implementable with the above mentioned prototype FIs. In this configuration the device is capable to provide intrinsic temperature noise down to $\sim 35$nKHz$^{-1/2}$  around 1K for a sufficiently large $h_{exc}$.  The proposed superconducting hybrid structure has the potential for the realization of effective on-demand on-chip temperature-to-frequency converters as well as ultrasensitive electron thermometers or radiation sensors easily integrable with current superconducting electronics.
\acknowledgements

We acknowledge J. S. Moodera and J. W. A. Robinson for fruitful comments.
F.G. acknowledges the European Research Council under the European Union's Seventh Framework Program (FP7/2007-2013)/ERC Grant agreement No. 615187-COMANCHE for funding. F.G. and  P.S. acknowledge MIUR-FIRB2013 -- Project Coca (Grant No.~RBFR1379UX) for partial financial support. P.S. has received funding from the European Union FP7/2007-2013 under REA Grant agreement No. 630925 -- COHEAT. A.B. thanks the support of the MIUR-FIRB2012 - Project HybridNanoDev (Grant No.RBFR1236VV). The work of F.S.B was supported by the Spanish Ministerio de Econom'a y Competitividad (MINECO) through the Project No. FIS2014-55987-P and Grupos Consolidados UPV/EHU del Gobierno Vasco (Grant No. IT-756-13).

\end{document}